# Edge States of α-Bismuthene Nanostructures

Sara Salehitaleghani[1], Tobias Maerkl[1], Pawel J Kowalczyk[2], Maxime Le Ster[2], Xiaoxiong Wang[3], Guang Bian[4], Tai-Chang Chiang[5], and Simon A Brown[1]*

[1] The MacDiarmid Institute for Advanced Materials and Nanotechnology, School of Physical and Chemical Sciences, University of Canterbury, Christchurch, New Zealand
[2] Department of Solid-State Physics, Faculty of Physics and Applied Informatics, University of Lodz, Lodz, Poland
[3] College of Science, Nanjing University of Science and Technology, Nanjing 210094, People's Republic of China
[4] Department of Physics and Astronomy, University of Missouri, Columbia, MO 65211, United States of America
[5] Department of Physics, University of Illinois at Urbana-Champaign, 1110 West Green Street, Urbana, IL 61801-3080, United States of America

E-mail: simon.brown@canterbury.ac.nz

## Abstract

We present a systematic investigation of the edge states of two-dimensional α-bismuthene (α-Bi) structures self-assembled on HOPG substrates, using scanning tunnelling microscopy and scanning tunnelling spectroscopy. The measurements are carried out for 3ML, 5ML and 7ML thick Bi structures. Our spectroscopy studies reveal clear features at the edges of the 5ML and 7ML thick structures, and the positions of the edge states (ESs) coincide with the topographical step edges. In contrast, in 3ML structures the ESs appear to be absent and instead new states are sometimes observed, far from the topographical edge. These states are associated with a moiré pattern and result from strain-induced modulation of the topology. Our observations demonstrate the impact on the edge states of coupling to adjacent structures.

## 1. Introduction

α-bismuthene (α-Bi) is a two-dimensional black-phosphorus-like allotrope of bismuth that is expected to have robust quantum spin Hall states [1], allowing spintronic applications such as long-distance spin transport and spin-to-charge conversion [2]. The SOC-induced bulk band gap of α-Bi [3] is large enough to make it a candidate for topotronic devices that operate efficiently at room temperature [2]. These properties originate from the band structure, which has a complex dependence on thickness that leads to interesting quantum size effects [4], and which derives from the non-symmorphic symmetry of the lattice (see Figure 1 (a, b)) [5].

Both α-Bi and the other (hexagonal) 2D allotrope of bismuth (β-Bi) are theoretically expected to have spin-polarized 1D edge states (ESs) [6-8]. α-Bi is energetically favoured at small thicknesses but β-Bi is generally observed at large thicknesses [9]. β-Bi structures have been thoroughly investigated [10-18], and it is found the nature of the ESs varies depending on the substrate on which it is grown. α-Bi structures were grown on several substrates (Si(111) [19], Ge(111) [20], the superconductor $NbSe_2$ [21], graphene supported by SiC [22, 23], $MoS_2$ [24], $TiSe_2$ [25] and HOPG [3, 26]), but less is known about their topological properties and ESs. Calculations suggest a strong dependence of the topology on the atomic buckling at the α-Bi surface i.e. the difference in height, $h$, between the surface atoms (compare Figure 1 (c) and (d)) [3]. It was reported that for *flat* surfaces ($h = 0$, Figure 1 (d)) the ESs are topologically protected and that a nontrivial-trivial transition occurs as the buckling is increased to $h \sim 0.1$ Å [3].

Here we report a comprehensive characterization of the edge states of α-Bi structures grown on HOPG, with emphasis on how the edges are affected by adjacent structures (either other Bi structures or the HOPG substrate). ESs for 5ML and 7ML structures are similar to those reported in ref. [3], including a characteristic peak at energies ~0.1 eV. However, we find no evidence for ESs on 3ML structures. Instead we observe states with similar characteristics to the previously reported ESs in along lines parallel to the fringes of a moiré pattern, and localised regions that are most likely associated with defects. These results indicate that the buckling of the α-Bi islands is modified locally, consistent with a topological transition.

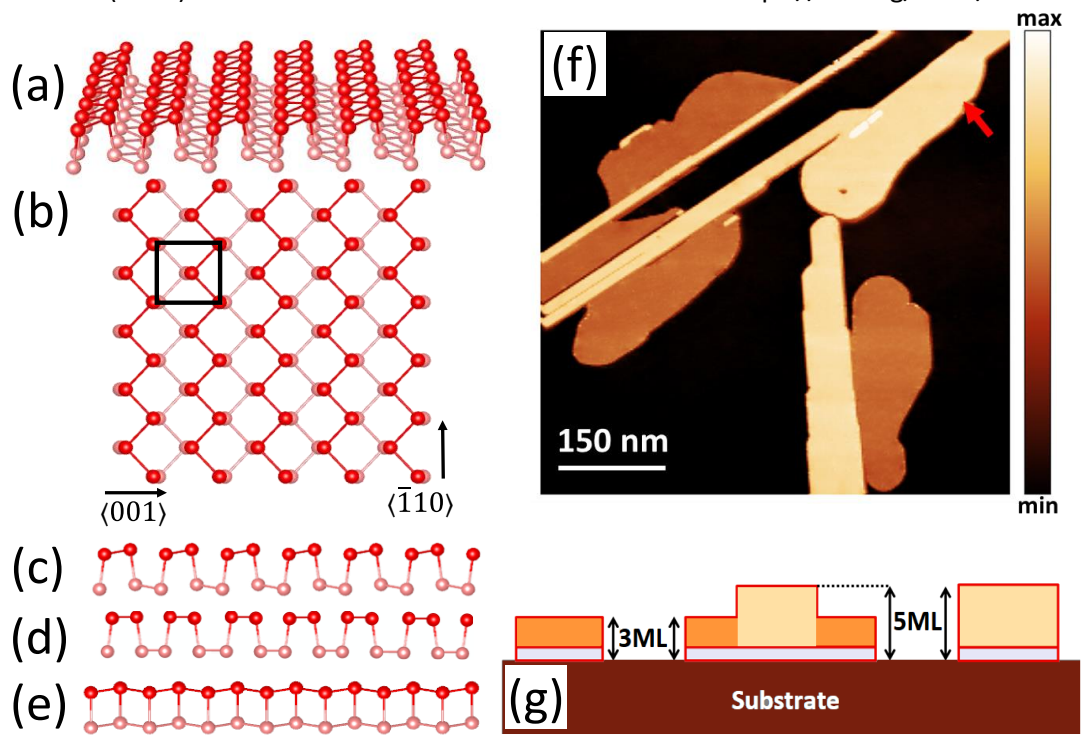

Figure 1 Black phosphorus structure and experimentally observed morphology for α-Bi. (a) Perspective, (b) top view, (c) side view along ⟨001⟩ for the buckled structure ($h \sim 0.1$ Å [3]), (d) side view along ⟨$\bar{1}$10⟩ for the structure with zero buckling, and (e) side view along ⟨001⟩ directions. (f) Large area STM image showing typical α-Bi island morphology when grown on HOPG. The islands typically comprise 3ML thick bases (darker orange) topped by additional 2ML stripes forming 5ML Bi and 7ML Bi (yellow and white respectively). Isolated 5ML and 7ML structures i.e. the 5ML and 7ML islands which sit directly on the HOPG substrate can occasionally occur (an example of a 5ML island is arrowed). (g) Schematic illustration of the basic morphology. Left: 3ML island. Centre: 5ML stripe (edge adjacent to 3ML island). Right: 5ML island (edge adjacent to the HOPG substrate). Colours are chosen to match the image in (f). Note that island edges are always adjacent to the substrate, whereas edges of stripes are adjacent to other α-Bi layers. Note also that the schematic includes an inert wetting layer which means that the observed density of states for each of the 3ML, 5ML and 7ML islands correspond to calculated band structures for 2ML, 4ML and 6ML α-Bi respectively [4].

## 2. Methods

1.2ML 5N purity Bi from a Knudsen cell was thermally deposited on freshly cleaved highly oriented pyrolytic graphite (HOPG) substrates at room temperature and under ultrahigh vacuum conditions (UHV, $<1 \times 10^{-9}$ mbar). The deposition rate was kept at ~0.01 Å s$^{-1}$ and black-phosphorus-like α-Bi structures are always observed for mean thicknesses less than ~12ML [27]. Scanning tunnelling microscopy (STM) and spectroscopy (STS) measurements were carried out using a variable-temperature UHV scanning tunnelling microscope (Scienta Omicron VT-AFM XA) system. The STS data was acquired in various bias voltage ranges (typically in the range ±1V with a setpoint current of 300 pA) and using a lock-in amplifier; modulation voltage and frequency were 20 mV and 2.7 kHz. All spectra for bulk α-Bi structures are in reasonable agreement with those measured previously using standard DC bias techniques [4]. All data reported here was obtained at low temperatures ~50 K.

## 3. Experimental Results

α-Bi typically forms [4, 5, 24, 26, 28-32] atomically flat structures with layered "wedding cake" profiles. Figure 1 (a-e) show the atomic structure and Figure 1 (f, g) show the typical morphology of the islands. Detailed STM measurements [28] show that the islands comprise 3ML thick bases (see Figure 1 and caption) topped by additional 2ML stripes, so that the stripes have total thicknesses of 5ML and 7ML. It is also occasionally possible to find isolated 5ML and 7ML islands which are not surrounded by other Bi structures. This allows us to investigate the impact of the interactions with adjacent structures on the nature of the ESs i.e. to compare results from edges adjacent to another Bi structure (e.g. 5ML stripe next to a 3ML base) and edges adjacent to HOPG (e.g. 5ML island).

Previous calculations show that the band structures change with the thickness of the material [4, 5] and so it is expected that the edge states and topology could vary as well. 3ML, 5ML and 7 ML structures exhibit very different STS spectra, corresponding to the results of calculations for freestanding 2ML, 4ML and 6ML α-Bi. (The difference in measured height is believed to be due to the presence of an inert wetting layer [4]). Therefore, we present data for structures of each thickness separately, starting by considering examples of ESs for 7ML structures (both islands and stripes); we then move on to a discussion of 5ML and 3ML structures.

### 3.1. 7ML Structures

A typical isolated 7ML Bi island is shown in Figure 2 (a). STS measurements were performed on the different edges of the island with magnified views shown in Figure 2 (b) to (e) and corresponding maps of the local density of states (LDOS) at +100 mV in Figure 2 (f) to (i). The edge states are immediately visible as prominent bright features (high-LDOS near the Fermi level; see also Figure S1). ESs are observed over the entire perimeter of the island, including the rounded ends.

This particular 7ML island exhibits some unusual additional complexity that provides further information. Firstly, the grain boundary arrowed in Figure 2 (g) separates two grains of α-Bi that are rotated by ~90° with respect to each other [29, 30] and therefore have different orientations with respect to the HOPG substrate. The observed ESs are very similar, which is a strong indication that the ESs are not sensitive to the orientation of Bi on HOPG. Secondly, this island happened to grow over an HOPG step edge and so the interaction of lower part of the island (Figure 2 (d)) with the substrate is expected to be different to that of the upper part of the island (Figure 2 (b,c)): the signature of the edge state remains unchanged, again suggesting the ESs are not affected by interactions with the substrate. The lack of dependence on the crystallographic orientation of the edge or interaction with the substrate indicates the ESs are inherent to the edge, are robust, and therefore are likely have a topological origin. [3]

Figure 3 (a) shows a 7ML Bi stripe that is surrounded by a 5ML Bi island and Figure 3 (b) shows a magnified view of the edge. The ES is consistently observed in the LDOS maps at energies ~+100 mV (see Figure 3 (c) and discussion of STS spectra below) and it is not dependent on the crystallographic orientation of the edges (see Figure S1). Despite the fact that the ES is adjacent to a material with a very different band structure (5ML Bi versus HOPG) the results for the ESs are qualitatively similar to those of the isolated 7ML Bi islands. Figure 3 (d) shows an LDOS intensity profile perpendicular to the edge of the stripe (along the white line in Figure 3 (c)). The ES is clearly resolved as an additional light blue feature at ~0.1 eV (between the red dashed lines that mark the position of the topographic edge). The ES decays completely into the bulk of the 7ML structure. Hence the ES is aligned with the topographical edge shown in the height profile in Figure 3 (e). To illustrate this more clearly Figure 3 (f) shows a mapping of the LDOS at +100 mV (Figure 3 (c)) onto the 3D topography image (Figure 3 (b)).

The intrinsic width of the ESs in our experiments cannot be exactly determined due to the tip broadening effect (see Figure S2). While the measured widths are close to those calculated for 2ML α-Bi nanoribbons (i.e. ~2 nm) [3], we believe that this agreement is accidental as the width is also found to correspond to the measured (i.e. tip-broadened) width of the topographical edge. The width is found to be slightly smaller for the stripes than for islands, simply because tip broadening is more significant for taller structures. Within the range that it can be observed, the position and width of the ES do not change as a function of bias (Figure S3).

The fact that the width of the ESs always corresponds to the measured width of the topographic step suggests that the ESs are physically located right at the topographic edge. Importantly, no enhancement of the DOS is observed on the flat part of the island – Figure 3 (f) shows clearly that the position of the ES corresponds to that shown schematically in Figure S2 (c). This is in strong contrast with previous calculations for α-Bi (see especially figure 4 in ref. [3]) which predicted that the ESs should extend *from* the topographic edge *into* the bulk, as illustrated schematically in Figure S2 (b). We emphasise that similar data were obtained for many islands and stripes of different thicknesses, and so the position of the ES is significant. Similar positioning of the ES was previously reported in calculations of the edge state band structure for monolayer β-Bi on α-Bi [11]: there the ES wavefunction clearly extends beyond the final atom in the overlayer in some cases and the differences in coupling lead to a modulation of the topology of the ESs.

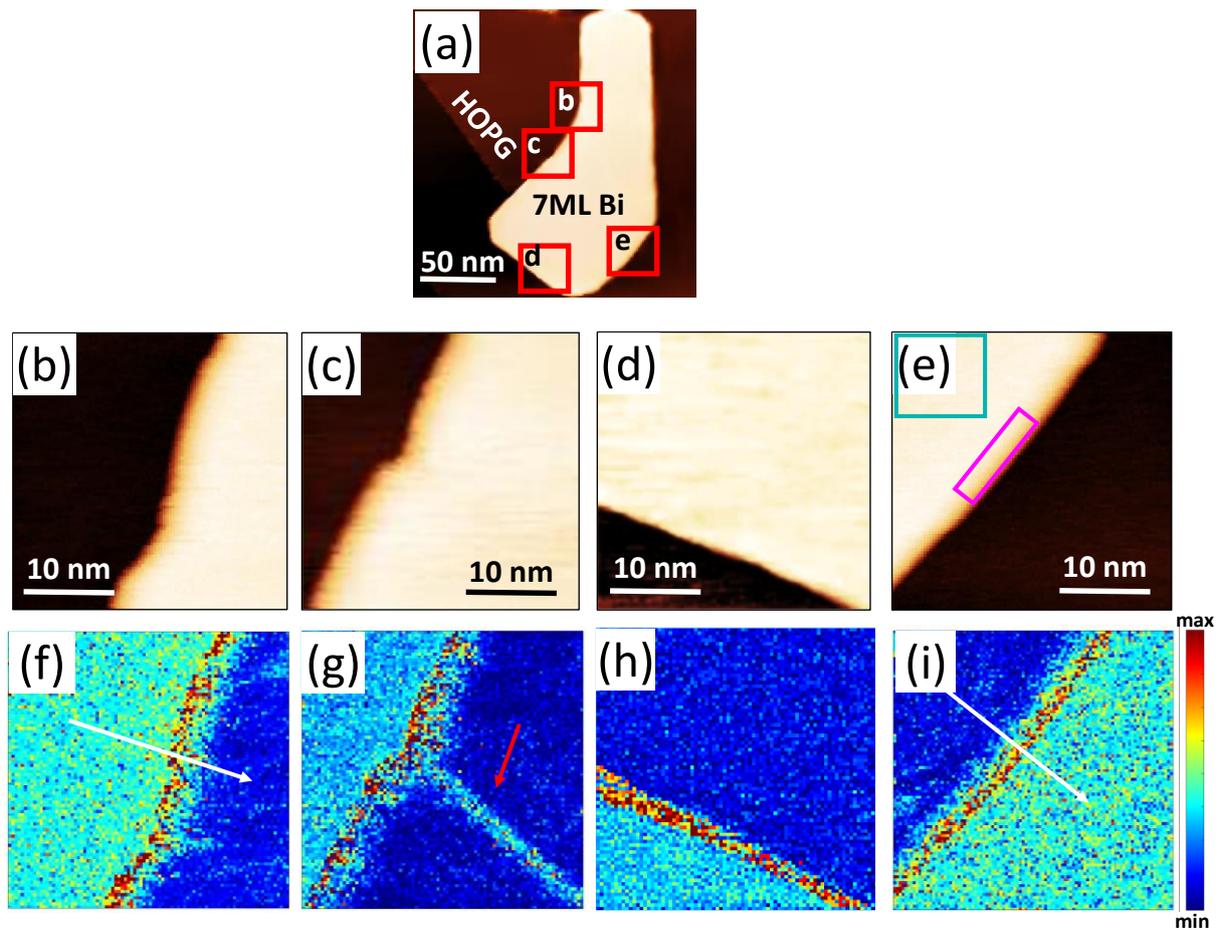

Figure 2 (a) Large-scale STM image of an isolated 7ML Bi island. Imaging conditions: $V_t = +1$ V, $I_t = 20$ pA and T = 50 K. (b)-(e) Magnified view of the regions inside the red squares in (a) showing various edges of the island with different crystallographic directions. The coloured rectangles in (e) correspond to spectra shown in Figure 4 (a). (f)-(i) Corresponding dI/dV(V) maps at +140 mV. STS parameters: ±1 V, 300 pA and 50 K. Note (i) a feature with similar characteristics to the ESs is observed at a grain boundary (arrowed in g)), (ii) even though the island grew over an HOPG step edge the signature of the edge state remains unchanged in (h) indicating that the interaction of the ES with the substrate is the same. DOS colour scale: red is high, blue is low.

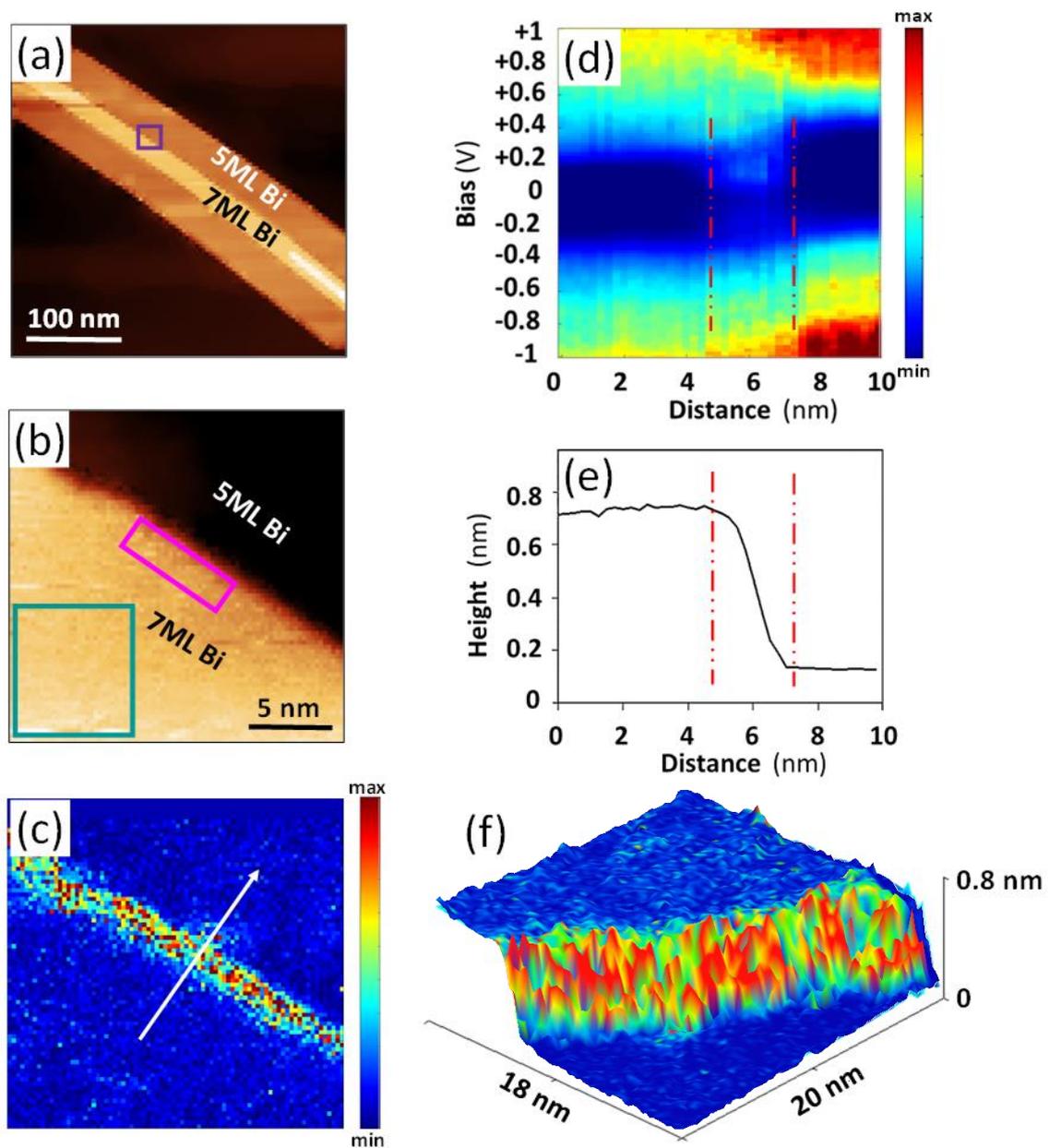

Figure 3 (a) STM image of a 7ML Bi stripe (adjacent to a 5ML Bi island). Imaging conditions: $V_t = +1$ V, $I_t = 20$ pA and T = 50 K. (b) Magnified view of the region inside the purple rectangle in (a) showing the edge. The coloured rectangles correspond to spectra shown in Figure 4 (b). (c) Corresponding LDOS map at +100 mV. (d) LDOS intensity profile drawn along the white arrow in (c), perpendicular to the edge of the 7ML Bi stripe. (e) Corresponding height profile. (f) Mapping of the +100 mV LDOS shown in (c) onto the STM 3D-topograph of (b); red regions exhibit the highest DOS and highlight that the ES is clearly located at the sloping edge of the stripe. STS parameters: ±1 V, 300 pA and 50 K.

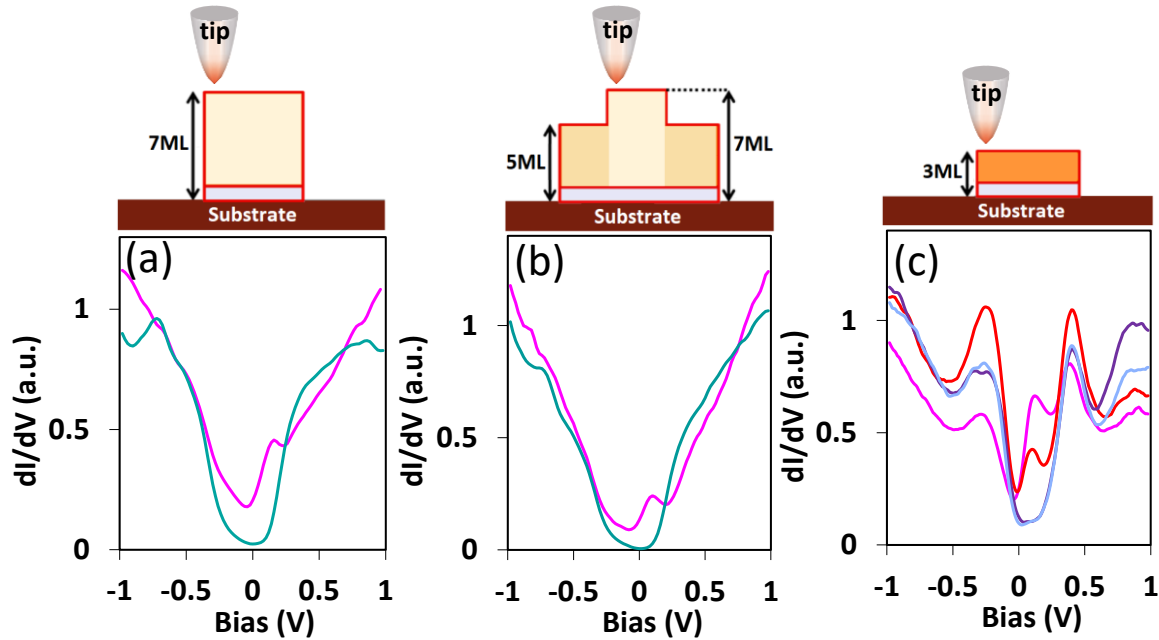

Figure 4  Experimental dI/dV(V) spectra from (a) an isolated 7ML Bi island and (b) a 7ML Bi stripe adjacent to a 5ML Bi structure. [Cartoons above the spectra illustrate the specific edges that are under investigation; colours correspond to the experimental image in Figure 1(g).] The spectra are extracted from the regions in the bulk (cyan) and at the edges (pink) of the structures inside the colour-coded rectangles in Figure 2 (e) and Figure 3 (b). (c) Comparison with spectra from a 3ML Bi island, which has a very different band structure but exhibits a similar peak at ~+150 mV – see Section 3.3. The spectra were recorded at different positions of the island marked by the colour-coded rectangles in Figure 6 (b). STS parameters: ±1 V, 300 pA and 50 K

The dI/dV(V) spectra obtained from a 7ML Bi island and from a 7ML stripe are shown in Figure 4 (a) and (b) respectively. The spectra were recorded at positions marked by the colour-coded rectangles in Figure 2 (e) and Figure 3 (b). The cyan curves in Figure 4 (a) and (b) are spectra obtained from the bulk (i.e. interior) of the stripe and the island respectively, and share strong similarities such as the shape of their LDOS valleys and the number and position of the main peaks [4]. Spectra from the edges of the two structures (pink curves) also have very similar shapes. They exhibit LDOS peaks at ~+100 mV that are absent in the spectra of bulk and thus are clearly associated with the ESs. It is important to note that the enhancement of the LDOS can be observed over an energy window between -200 mV and +250 mV. Hence these observations are consistent with the previous calculations [3] suggesting that the ES dispersion spans an energy range corresponding to the energy gap of the Bi. Note that near $E_F$ the LDOS at the edges is always higher than the corresponding bulk. The higher energy peaks in the spectra (particularly those at +0.3 eV and -0.4 eV) observed in the bulk of 7ML Bi appear to be suppressed at the edges, suggesting that those states are involved in the formation of the ESs.

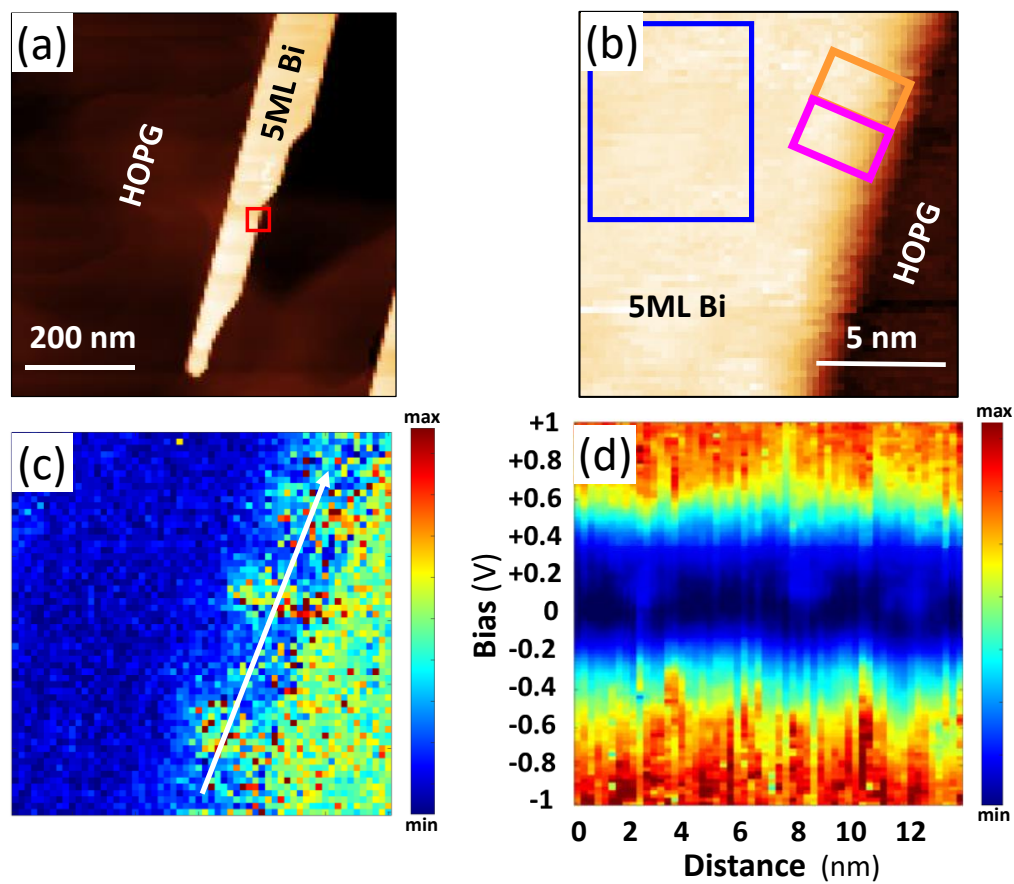

Figure 5 (a) STM image of an isolated 5ML Bi island. (b) Magnified view of the region inside the red rectangle in (a) showing the edge. The coloured rectangles correspond to spectra shown in Figure S5 (a). (c) +150 mV LDOS map corresponding to (b). (d) LDOS intensity profile drawn parallel to the edge along the white arrow in (c). Imaging conditions: $V_t = +1$ V, $I_t = 20$ pA and T = 50K. STS parameters: ±1 V, 300 pA and 50 K.

*3.2. 5ML Structures*

Results obtained for 5ML Bi stripes are similar to those of the 7ML Bi stripes i.e. continuous ESs are visible at ~+100 mV, which coincide with the topographical edge (Figure S4). The ESs have a width of ~3nm, as shown in Figure S6 (c) and (d). For isolated 5ML Bi islands, however, the ESs appear rather differently. Figure 5 (a) and (b) show a large-scale image and a magnified view respectively, while Figure 5 (c) and (c) show the corresponding LDOS map at +150 mV and a LDOS intensity profile drawn parallel to the edge (d). The key feature in Figure 5(c) is that the ES near $E_F$ obviously has a modulated intensity. The modulation of the ES is clearly seen in spectra (Figure S5) but the modulation does not seem to be periodic even in wider-scale images (not shown). This aperiodic behaviour may be partially explained by previously published [23] atomic resolution images which show that the reconstruction of Bi edges is not fully periodic. It is possible that the ESs could be better understood through an investigation of the reconstruction of dangling bonds at the edges, which might be expected to be different for the edges of islands (where five atomic layers terminate) and edges of stripes (where only two atomic layers terminate). However, we were unable to obtain reliable atomic scale images of the edges; our experimental conditions, including the tips, are optimised for the long scans required for spectroscopy, and typically the tips that provide the best spectra are not sharp enough to obtain good atomic resolutions images. To our knowledge only one example of an atomic resolution image of the edge reconstruction in α-Bi has been reported in the literature [23]. We note however that discontinuous ESs have previously been observed for WTe$_2$ systems due to structural inhomogeneities [33-35]. Another possibility is that the edge states couple differently to the different adjacent structures (HOPG or Bi).

### *3.3. 3ML Structures*

An STM image of a 3ML Bi island is shown in Figure 6 (a) along with magnified views of its edges in Figure 6 (b-d). The corresponding LDOS maps at +150 mV are shown in Figure 6 (e-g). No obvious electronic state is observed at the topographic edges, and in fact we were unable to observe ESs for any 3ML islands, despite making measurements under a range of imaging conditions using both lock-in and standard DC bias techniques. This is very much in contrast with Ref. [3] where ESs were observed for 3ML structures.

Interestingly, linear features with a high LDOS at ~+150 mV are observed ~5 nm away from the edge in Figure 6 (e) and (f) (white arrows). Isolated regions with high LDOS are also observed at various positions (red arrows in Figure 6 (e)) and turn out to have similar spectral characteristics. The spectra measured on the high LDOS regions (pink and red curves in Figure 4 (c), corresponding to the colour-coded rectangles in Figure 6 (b)) exhibit a clear peak at ~+150 mV, which is completely absent in spectra taken from other positions of the island (purple and blue curves in Figure 4 (c)).

Figure 6 (g) provides an important indication as to the origin of these linear features int eh 3ML islands: it shows a *periodic* modulation of the LDOS intensity *perpendicular* to the edge. This modulation is consistent with the presence of a moiré pattern (MP) [36]. A similar modulation is shown in Figure S7 (c) over a larger spatial region for a different island. The orientation of the MP (parallel to the edges of the island) is unusual for Bi on HOPG [37], but detailed analysis (Figure S8 (a)) reveals that this type of the MP is in fact expected when the Bi ⟨$\bar{1}$10⟩ direction is rotated~28° from the HOPG ⟨10$\bar{1}$0⟩ direction.

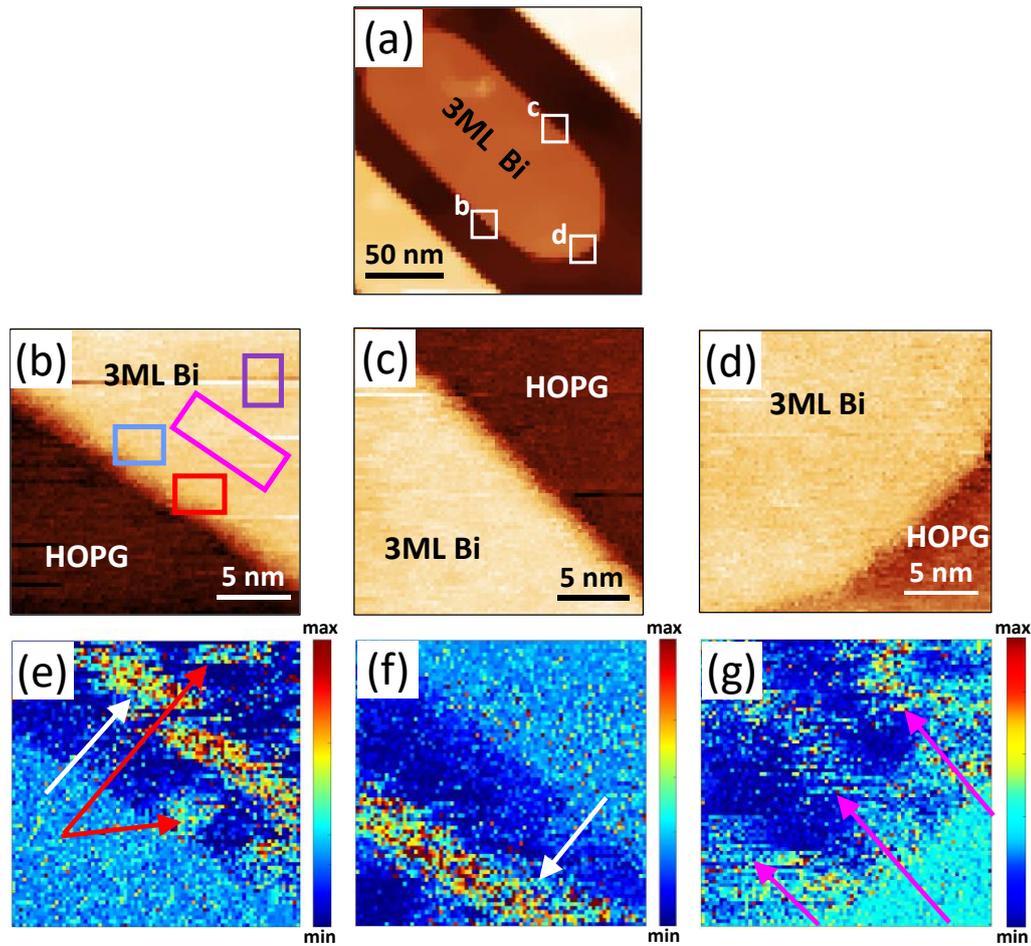

Figure 6 (a) STM image of a 3ML Bi island on HOPG. (b-d) Magnified views of the regions inside the white rectangles in (a) showing different edges. The coloured rectangles in (b) correspond to spectra shown in Figure 4 (c). Imaging conditions: $V_t = +1$ V, $I_t = 20$ pA and $T = 50$ K. (e-g) +150mV LDOS maps corresponding to (b-d) showing that no clear edge state is observed on 3ML structures; white arrows indicate an additional extended bright LDOS feature near the edges and red arrows (in (e)) indicate localised bright regions inside the bulk and at the edges. Pink arrows in (g) indicate LDOS features associated with moiré fringes in the bulk of the island. STS parameters: ±0.5 V, 300 pA and 50 K.

The MP causes a modulation of the strain in the α-Bi structure [30] and we believe that this induces different amounts of buckling, and hence different topologies, in each region. The position of the peak in the LDOS at ~+150 mV agrees well with that of the topologically protected edge bands calculated previously in Ref. [3] for freestanding 2ML α-Bi nanoribbon, and with the results presented above for 5ML and 7ML structures. Hence it is possible that the additional features (both the high LDOS 1D stripes and localised patches arrowed in Figure 6 (e-g)) might be associated with topological states. Modulation of topology by a moiré pattern has previously been predicted only in a very few special cases [38, 39] and experimentally observed in β-Bi/α-Bi heterostructures [11]. In this interpretation the crests and troughs of the MP would each correspond to different topological states, as predicted in [38]. Other interpretations are however possible, and for example it is possible that the observed stripes correspond to the *edges* of linear topological regions created by the MP [39]. In that interpretation, the MP must have twice the period observed in Figure 6 (g); Figure S8 (b) shows that such a MP is in fact possible if the Bi lattice is allowed to expand fractionally.

## 4. Summary and Discussion

We used scanning tunnelling spectroscopy at ~50 K to reveal edge states for 7ML and 5ML Bi nanostructures on HOPG substrate. ESs are observed for both islands (i.e. 5ML and 7ML Bi adjacent to bare HOPG substrates) and stripes (i.e. 5ML Bi structures adjacent to 3ML, and 7ML structures adjacent to 5ML). High LDOS features are only visible near the Fermi level, and are always located at the topographic edges, and not on the flat part of the islands as expected from previous calculations [3] ESs are found on the whole perimeter of the 7ML and 5ML Bi nanostructures including their round ends. This insensitivity to the different crystallographic directions of the edges provides a strong argument that the ES do not originate simply from a reconstruction (which would be expected to be different on edges with different crystallographic facets), and might be indicative of the topological origin of ESs, as previously reported by Lu et al. [3]. In the case of 5ML islands, the intensity of the ESs is modulated, whereas no modulation was observed for the 5ML stripes. This appears to be because the topological states couple differently to the adjacent structures and/or that the edge reconstruction is different for stripes and islands.

We found no clear signature of ESs for 3ML Bi islands. Instead, we observed localised states (high conductance regions at ~+150 mV) in small regions near the edges, and in extended regions along the fringes of a moiré pattern. We suggest that these states most likely result from a modulation of the local strain (by defects and by the moiré pattern) [40]. The localised states certainly do not originate from the reconstruction of dangling bonds since they are found far from the edges. Such a modulation of the topology could be valuable for engineering devices [38, 39].

The absence of previously reported [3] ESs for 3ML Bi islands and the presence of localised states with ES-like spectra can more generally be understood in terms of the sensitivity of the topology to the level of buckling in the α-Bi structure [5]. The buckling is likely to be controlled by local interaction with the underlying HOPG, the Bi/HOPG orientation, electronic coupling to the substrate, and the presence of grain boundaries (as well as strain and defects, as mentioned in the previous paragraph; note also the appearance of an ES-like feature in Figure 2 (c) at a grain boundary, which can be regarded as a chain of defects). Previous μLEED patterns for the 3ML structures exhibit weak spots that are consistent with a small amount of buckling (or a buckling that is modulated across the island), but these spots are completely absent in thicker structures, consistent with an absence of buckling and non-symmorphic symmetry of the lattice [4]. It also seems likely that electronic states in the 3ML structures are more strongly influenced than thicker structures by coupling to the substrate and charge transfer, and this might account for a higher sensitivity of 3ML ESs to perturbations. New calculations are required to fully understand the topological states in the hybrid α-Bi/HOPG system, however such calculations are very challenging due to the lack of commensuration between the α-Bi and HOPG lattices (very large supercells are required) and the inter-relationships between the effects of buckling, strain, substrate coupling, and the long-range periodicity of the moiré pattern.

## 5. Acknowledgement

This work was supported by the MacDiarmid Institute for Advanced Materials and Nanotechnology and the Marsden Fund (SS, TM, MLS, SAB), the National Science Centre, Poland (DEC-2015/17/B/ST3/02362, PJK), the National Natural Science Foundation of China (11204133, XXW) and the U.S. National Science Foundation (NSF-DMR-1809160) and Department of Energy, Office of Science, Office of Basic Energy Sciences, Division of Materials Science and Engineering (DE-FG02-07ER46383)